\documentstyle[11pt]{article}
\newcommand{\reff}[1]{eq.~(\ref{#1})}

\newcommand{\beq}{\begin{equation}}
\newcommand{\eeq}{\end{equation}}
\newcommand{\bea}{\begin{eqnarray}}
\newcommand{\eea}{\end{eqnarray}}
\newcommand{\bq}{\begin{quote}}
\newcommand{\eq}{\end{quote}}

\def\a{\alpha}
\def\b{\beta}

\def\e{\epsilon}
\def\g{\gamma}

\def\l{\lambda}

\def\D{\Delta}
\def\L{\Lambda}

\def\beq{\begin{equation}}
\def\eeq{\end{equation}}
\def\bea{\begin{eqnarray}}
\def\eea{\end{eqnarray}}
\def\ba{\begin{array}}
\def\ea{\end{array}}

\begin{document}
\begin{titlepage}
\begin{flushright}
KCL-TH-94-20\\
UQMATH-94-10\\
hep-th/9411241
\end{flushright}
\vskip.3in
\begin{center}
{\huge Solutions of the Yang-Baxter Equation with Extra
Non-Additive Parameters II: $U_q(gl(m|n))$}
\vskip.3in
{\Large Gustav W. Delius}
\footnote{Supported by Habilitationsstipendium der Deutschen
Forschungsgemeinschaft.}
\footnote{On leave from Department of Physics, Bielefeld University,
Germany.}
\vskip.1in
{\large Department of Mathematics, King's College London, Strand,
London WC2R 2LS, UK}
email: delius@mth.kcl.ac.uk
\vskip.2in
{\Large Mark D. Gould, Jon R. Links} and {\Large Yao-Zhong Zhang}
\footnote{Address after Feb. 1, 1995: Yukawa Institute for Theoretical
Physics, Kyoto University, Japan.}
\vskip.1in
{\large Department of Mathematics, University of Queensland, Brisbane,
Qld 4072, Australia}
email: yzz@maths.uq.oz.au
\end{center}
\vskip.6in
\begin{center}
{\bf Abstract:}
\end{center}
The type-I quantum superalgebras are known to admit
non-trivial one-parameter families of inequivalent finite dimensional irreps,
even for generic $q$. We apply the recently developed technique to
construct new solutions to the quantum Yang-Baxter equation associated
with the one-parameter family
of irreps of $U_q(gl(m|n))$, thus obtaining R-matrices which depend
not only on a spectral parameter but
in addition on further continuous parameters. These extra parameters
enter the Yang-Baxter equation in a similar way to the spectral parameter
but in a non-additive form.


\end{titlepage}
\newpage

In \cite{DGZ}, we developed a systematic method for constructing
R-matrices (solutions of the quantum Yang-Baxter equation (QYBE))
associated with the multiplicity-free tensor product of
{\em any} two affinizable irreps of
a quantum algebra. This approach was applied and extended to quantum
superalgebras in \cite{Bra94a,Del94b}. For the type-I quantum superalgebra
$U_q(gl(m|1))$ in particular, we were able to obtain R-matrices
depending continuously on extra parameters, entering in a similar way
as the spectral parameter but in a nonadditive form.
In this paper, we continue this study
to construct new such R-matrices for the type-I
quantum superalgebra $U_q(gl(m|n))$ for any $m\geq n$.

The freedom of having extra continuous parameters in R-matrices opens up new
and exciting possibilities. For example, in \cite{Bra94b}, by using the
R-matrix associated with the one-parameter family of 4-dimensional
irreps of $U_q(gl(2|1))$, we derived a new exactly solvable
lattice model of strongly correlated electrons on the unrestricted
$4^L$-dimensional electronic Hilbert space $\otimes^L_{n=1}{\bf C}^4$
(where $L$ is the lattice length), which is a $gl(2|1)$
supersymmetric generalization of the Hubbard model with the Hubbard
on-site interaction coupling coefficient related to the parameter carried by
the 4-dimensional irrep.

The origin of the extra parameters in our solutions are the
parameters which are carried by the irreps themselves of the associated
quantum superalgebra.
As is well known \cite{Kac77},  type-I superalgebras admit nontrivial
one-parameter families of finite-dimensional irreps which
deform to provide
one-parameter families of finite-dimensional irreps of the corresponding
type-I quantum superalgebras, for generic $q$ \cite{MS93}.
Note however that for quantum simple bosonic Lie algebras families of
finite-dimensional
representations are possible only when the deformation parameter $q$ is
a root of unity. Therefore our solutions are not related to the chiral
Potts model R-matrices which arises from quantum bosonic algebras at $q$ a root
of unity only \cite{Baz90,Dat91}.

Let us have a brief review of our general formalism formulated in
\cite{Bra94a,Del94b}.
Let $G$ denote a simple Lie superalgebra of rank $r$ with generators
$\{e_i, f_i, h_i\}$ and let $\alpha_i$ be its simple roots. Then the quantum
superalgebra $U_q(G)$ can be defined with the structure of a
${\bf Z}_2$-graded quasi-triangular Hopf algebra.
We will not give the full
defining relations of $U_q(G)$ here but mention that $U_q(G)$ has a
coproduct structure given by
\begin{equation}
\Delta(q^{h_i/2})=q^{h_i/2}\otimes q^{h_i/2}\,,~~~\Delta(a)=a\otimes
  q^{-h_i/2}+q^{h_i/2}\otimes a\,,~~~a=e_i, f_i.
\end{equation}
The multiplication rule for the tensor product is defined for elements
$a,b,c,d\in U_q(G)$ by
\begin{equation}\label{gradprod}
(a\otimes b)(c\otimes d)=(-1)^{[b][c]}(ac\otimes bd)
\end{equation}
where $[a]\in {\bf Z}_2$ denotes the degree of the element $a$.

Let $\pi_\a$ be a one-parameter family of irreps of $U_q(G)$ afforded by
the irreducible module $V(\L_\a)$ in such a way that the highest weight of
the irrep depends on the parameter $\a$.
Assume for any $\a$ that the irrep $\pi_\a$
is affinizable, i.e. it can be extended to an irrep of the corresponding
quantum affine superalgebra $U_q(\hat{G})$. Consider an operator
(R-matrix) $R(x|\a,\b)\in {\rm End}(V(\L_\a)\otimes V(\L_\b))$,
where $x\in {\bf C}$ is the usual spectral parameter and
$\pi_{\a},~\pi_{\b}$ are two irreps from the one-parameter
family. It has been
shown by Jimbo \cite{Jimbo} that a solution to the linear equations
\begin{eqnarray}
&&R(x|\a,\b)\Delta^{\a\b}(a)=\bar{\Delta}^{\a\b}
  (a)R(x|\a,\b)\,,~~~\forall a\in U_q(G),\nonumber\\
&&R(x|\a,\b)\left (x\pi_{\a}(e_0)\otimes \pi_{\b}(q^{-h_0/2})+
  \pi_\a (q^{h_0/2})\otimes \pi_{\b}(e_0)\right )\nonumber\\
&&~~~~~~  =\left (x\pi_{\a}(e_0)\otimes \pi_{\b}(q^{h_0/2})
  +\pi_{\a}(q^{-h_0/2})\otimes \pi_{\b}(e_0)\right )R
  (x|\a,\b)\label{r(x)1}
\end{eqnarray}
satisfies the QYBE
in the tensor product module $V(\L_\a)\otimes V(\L_\b)\otimes V(\L_\g)$
of three irreps from the one-parameter family:
\begin{equation}
R_{12}(x|\a,\b)R_{13}(xy|\a,\g)R_{23}(y|\b,\g)
  =R_{23}(y|\b,\g)R_{13}(xy|\a,\g)R_{12}(x|\a,\b).
\end{equation}
In the above,
$\bar{\Delta}=T\cdot \Delta$, with $T$ the twist map defined by
$T(a\otimes b)=(-1)^{[a][b]}b\otimes a\,,~\forall a,b\in U_q(G)$ and
$\Delta^{\a\b}(a)=(\pi_{\a}\otimes \pi_{\b})\Delta(a)$;
also, if $R(x|\a,\b)=\sum_i\pi_{\a}(a_i)\otimes
\pi_{\b}(b_i)$, then $R_{12}(x|\a,\b)=\sum_i\pi_{\a}(a_i)
\otimes\pi_{\b}(b_i)\otimes I$ etc. Jimbo also showed that the solution to
(\ref{r(x)1}) is unique, up to scalar functions. The multiplicative
spectral parameter $x$ can be transformed into an additive spectral
parameter $u$ by $x=\mbox{exp}(u)$.

In all our equations we implicitly use the ``graded" multiplication rule of
\reff{gradprod}. Thus the R-matrix of a quantum superalgebra satisfies
a ``graded" QYBE which, when written as an ordinary
matrix equation, contains extra signs:
\begin{eqnarray}
&&\left(R(x|\a,\b)\right)_{ij}^{i'j'}
\left(R(xy|\a,\g)\right)_{i'k}^{i''k'}
\left(R(y|\b,\g)\right)_{j'k'}^{j''k''}
(-1)^{[i][j]+[k][i']+[k'][j']}\nonumber\\
&&~~~~~=\left(R(y|\b,\g)\right)_{jk}^{j'k'}
\left(R(xy|\a,\g)\right)_{ik'}^{i'k''}
\left(R(x|\a,\b)\right)_{i'j'}^{i''j''}
(-1)^{[j][k]+[k'][i]+[j'][i']}.
\end{eqnarray}
However after a redefinition
\beq\label{redef}
\left(\tilde{R}(\cdot|\a,\b)\right)_{ij}^{i'j'}
=\left(R(\cdot|\a,\b)\right)_{ij}^{i'j'}\,
(-1)^{[i][j]}
\eeq
the signs disappear from the equation. Thus any solution of the ``graded"
QYBE arising from the R-matrix of a quantum superalgebra
provides also a solution of the standard QYBE after
the redefinition in \reff{redef}.

Introduce the graded permutation operator $P^{\a\b}$
on the tensor product
module $V(\L_\a)\otimes V(\L_\b)$ such that
\begin{equation}\label{gradperm}
P^{\a\b}(v_\alpha\otimes v_\beta)=(-1)^{[\alpha][\beta]}
  v_\beta\otimes v_\alpha\,,~~
  \forall v_\alpha\in V(\L_\a)\,,~v_\beta\in V(\L_\b)
\end{equation}
and set
\begin{equation}
\check{R}(x|\a,\b)=P^{\a\b}R(x|\a,\b).
\end{equation}
Then (\ref{r(x)1}) can be rewritten as
\begin{eqnarray}
&&\check{R}(x|\a,\b)\Delta^{\a\b}(a)=\Delta^{\b\a}(a)
  \check{R}(x|\a,\b)\,,~~~\forall a\in U_q(G),\nonumber\\
&&\check{R}(x|\a,\b)\left (x\pi_{\a}(e_0)\otimes\pi_{\b}
  (q^{-h_0/2})+\pi_{\a}(q^{h_0/2})\otimes
  \pi_{\b}(e_0)\right )\nonumber\\
&&~~~~~~  =\left (\pi_{\b}(e_0)\otimes \pi_{\a}(q^{-h_0/2})+
  x\pi_{\b}(q^{h_0/2})\otimes \pi_{\a}(e_0)\right )
  \check{R}(x|\a,\b)\label{r(x)2}
\end{eqnarray}
and in terms of $\check{R}(x|\a,\b)$ the QYBE becomes
\bea
&&(I\otimes\check{R}(x|\a,\b))(\check{R}(xy|\a,\g)
  \otimes I)(I\otimes\check{R}(y|\b,\g))\nonumber\\
&&~~~~~~=(\check{R}(y|\b,\g)\otimes I)(I\otimes \check{R}(xy|\a,\g))
  (\check{R}(x|\a,\b)\otimes I)\label{non-additive-ybe}
\eea
both sides of which act from $V(\L_\a)\otimes V(\L_\b)\otimes V(\L_\g)$
to $V(\L_\g)\otimes V(\L_\b)\otimes V(\L_\a)$. Note that this equation,
if written in matrix form, does not have extra signs.
This is because the definition of the graded permutation operator
in \reff{gradperm} includes the signs of \reff{redef}.
In the following we will normalize the $R$-matrix
$\check{R}(x|\a,\b)$ in such a way that
$\check{R}(x|\a,\b)\check{R}(x^{-1}|\b,\a)=I$,
which is usually called the unitarity condition in the literature.

For three special values of $x$:
$x=0,~x=\infty$ and $x=1$,
$\check{R}(\cdot|\a,\b)$ satisfies the spectral-free,
but extra non-additive-parameter-dependent QYBE,
\bea
&&(I\otimes\check{R}(\cdot|\a,\b))(\check{R}(\cdot|\a,\g)
  \otimes I)(I\otimes\check{R}(\cdot|\b,\g))\nonumber\\
&&~~~~~=(\check{R}(\cdot|\b,\g)\otimes I)(I\otimes \check{R}(\cdot|\a,\g)
  (\check{R}(\cdot|\a,\b)\otimes I).
\eea
In the case of a multiplicity-free tensor product decomposition
\begin{equation}
V(\L_\a)\otimes V(\L_\b)=\bigoplus_\mu V(\mu),\label{free}
\end{equation}
where $\mu$ denotes a highest weight depending on the parameters $\a$
and $\b$, the $\check{R}(0|\a,\b),~\check{R}(\infty|\a,\b)$ and
$\check{R}(1|\a,\b)$ may be obtained in the particularly simple form
\cite{DGZ,Bra94a,Del94b}
{}~\footnote{Similar relations to (\ref{r-0}) and (\ref{r-infty})
for quantum bosonic algebras were obtained in
\cite{Res88,Dri90}.}:
\begin{eqnarray}
&&\check{R}(0|\a,\b)=\sum_{\mu}
  \epsilon(\mu)\; q^{\frac{C(\mu)-C(\L_\a)-C(\L_\b)}{2}}\;
  {\bf P}^{\a\b}_\mu,\label{r-0}\\
&&\check{R}(\infty|\a,\b)=\sum_{\mu}
  \epsilon(\mu)\; q^{-\frac{C(\mu)-C(\L_\a)-C(\L_\b)}{2}}\;
  {\bf P}^{\a\b}_\mu,\label{r-infty}\\
&&\check{R}(1|\a,\b)=\sum_{\mu}\;
  {\bf P}^{\a\b}_\mu,\label{r-1}
\end{eqnarray}
where $C(\Lambda)=(\Lambda, \Lambda+2\rho)$ is the eigenvalue of the quadratic
Casimir invariant of $G$ in the irrep with highest weight $\Lambda$, $\rho$
is the graded half-sum of positive roots of $G$
and ${\bf P}^{\a\b}_\mu:~V(\L_\a)\otimes V(\L_\b)\rightarrow
V(\mu)\subset V(\L_\b)\otimes V(\L_\a)$ are the elementary intertwiners, i.e.,
${\bf P}^{\a\b}_\mu\Delta^{\a\b}(a)=\Delta^{\b\a}(a)
{\bf P}^{\a\b}_\mu$,
{}~ $ \forall a\in U_q(G)$;
$\epsilon(\mu)$ is the
parity of $V(\mu)$ in $V(\L_\a)\otimes V(\L_\b)$.
Since in the present case $\a$ etc are continuous parameters,
the parities $\epsilon(\mu)$ can easily be
worked out by examining the limit $\a\rightarrow\b$.

The elementary intertwiners satisfy the relations
\begin{eqnarray}
&&{\bf P}^{\a\b}_\mu\,{\cal P}^{\a\b}_{\mu'}=
  {\cal P}^{\b\a}_{\mu'}\,{\bf P}^{\a\b}_\mu=\delta_{\mu\mu'}
  {\bf P}^{\a\b}_\mu\,,\nonumber\\
&&{\bf P}^{\b\a}_\mu\,{\bf P}^{\a\b}_{\mu'}=\delta_{\mu\mu'}
  {\cal P}^{\a\b}_\mu
\eea
where the ${\cal P}^{\a\b}_\mu:~V(\L_\a)\otimes V(\L_\b)
\rightarrow V(\mu)\subset V(\L_\a)\otimes V(\L_\b)$ are projection operators
satisfying
\beq
{\cal P}^{\a\b}_\mu\,{\cal P}^{\a\b}_{\mu'}
  =\delta_{\nu\nu'}{\cal P}^{\a\b}_\nu\,,~~~~\sum_\mu
  {\cal P}^{\a\b}_\mu=I.\label{projectors}
\eeq
Let $\{|e^\mu_i\rangle_{\a\otimes \b}\}$ be an orthonormal
basis for $V(\mu)$ in $V(\L_\a)\otimes V(\L_\b)$.
$V(\mu)$ is also embedded in $V(\L_\b)\otimes V(\L_\a)$ through the
opposite coproduct $\bar{\Delta}$. Let
$\{|e^\mu_i\rangle_{\b\otimes \a}\}$ be the
corresponding orthonormal
basis \footnote{For the precise definition of this basis see Appendix
C of \cite{DGZ}.}.
Using these bases the operators ${\cal P}^{\a\b}_{\mu}$
and ${\bf P}^{\a\b}_{\mu}$ can be expressed as
\begin{eqnarray}
&&{\cal P}^{\a\b}_\mu=\sum_i \left |e^\mu_i\right
  \rangle_{\a\otimes
  \b~\a\otimes\b}\!\left\langle e^\mu_i\right |,\nonumber\\
&&{\bf P}^{\a\b}_\mu=\sum_i\left |e^\mu_i\right
  \rangle_{\b\otimes\a~
  \a\otimes\b}\!\left\langle e^\mu_i\right |\,.
\end{eqnarray}

The most general $\check{R}(x|\a,\b)$ satisfying the first
equation in (\ref{r(x)2}) may be written in the form
\begin{equation}
\check{R}(x|\a,\b)=\sum_{V(\mu)\in V(\L_\a)\otimes V(\L_\b)}
  \rho_\mu(x)\;{\bf P}^{\a\b}_\mu\label{ansatz}
\end{equation}
where $\rho_\mu(x)$, are  unknow functions depending on
$x,~q$ and the extra non-additive parameters $\a$ and $\b$.
It follows from the second equation of (\ref{r(x)2})
\cite{DGZ,Bra94a,Del94b} that if
\beq
{\cal P}^{\a\b}_\mu\left (
  \pi_{\a}(e_0)\otimes \pi_{\b}(q^{-h_0/2})\right )
  {\cal P}^{\a\b}_{\mu'}\neq 0\label{pp2}
\eeq
then the coefficients $\rho_\mu(x)$ in (\ref{ansatz}) are given
recursively by
\begin{equation}\label{2rho1}
\rho_\mu(x)=\rho_{\mu'}(x)\frac{q^{C(\mu)/2}
  + \epsilon(\mu)\epsilon(\mu') xq^{C(\mu')/2} }{xq^{C(\mu)/2}
  +\epsilon(\mu)\epsilon(\mu') q^{C(\mu')/2} },~~~~\forall \mu\neq\mu'.
\end{equation}
$\e(\mu)\e(\mu')=-1$ always if (\ref{pp2})
is satisfied. With the help of notation
\beq\label{angle}
\left\langle a\right\rangle\equiv \frac{1-xq^a}{x-q^a},
\eeq
(\ref{2rho1}) then becomes
\beq
\rho_\mu(x)=\left\langle\frac{C(\mu')-C(\mu)}{2}\right\rangle
  \rho_{\mu'}(x)\,.\label{2rho2}
\eeq

We have a relation between the coefficients
$\rho_\mu$ and $\rho_{\mu'}$ whenever the condition
\reff{pp2} is satisfied, i.e., whenever $\pi_\a(e_0)\otimes \pi_\b(q^{-h_0/2})$
maps from the module $V(\mu')$ to the module $V(\mu)$.
As a graphical aid \cite{Zha91} we introduce the tensor product graph.
\vskip.1in
\noindent {\bf Definition 1:}
The {\bf tensor product graph} (TPG)  associated to the tensor product
$V(\L_\a)\otimes V(\L_\b)$ is a graph
whose vertices are the irreducible modules
$V(\mu)$ appearing in the decomposition  of
$V(\L_\a)\otimes V(\L_\b)$. There is an edge between a vertex $V(\mu)$
and a vertex $V(\mu')$ iff
\beq\label{link}
{\cal P}^{\a\b}_\mu\left (\pi_\a(e_0)\otimes \pi_\b(q^{-h_0/2})\right )
  {\cal P}^{\a\b}_{\mu'}\neq 0.
\eeq
\vskip.1in
If $V(\L_\a)$ and $V(\L_\b)$ are irreducible $U_q(G)$-modules then the
TPG is always connected, i.e., every node is linked to every
other node by a path of edges.
This implies that the relations \reff{2rho2} are sufficient to
determine all the coefficients $\rho_\mu(x)$ uniquely, up to an overall
factor.
If the TPG is multiply connected, i.e., if there exist
more than two paths between two nodes, then the relations overdetermine
the coefficients, i.e., there are consistency conditions.
However, because the existence of a solution to the Jimbo equations
is guaranteed by
the existence of the universal R-matrix, these consistency conditions
will always be satisfied.

The straightforward but tedious and impractical way to determine the
TPG is to work out explicitly the left hand side  of (\ref{link}).
It is much more practical to work instead with the following larger graph
which is often enough to determine the coefficients $\rho_\mu(x)$.
\vskip.1in
\noindent {\bf Definition 2:}
The {\bf extended tensor product graph} (ETPG) associated to the
tensor product $V(\L_\a)\otimes V(\L_\b)$ is a graph
whose vertices are the irreducible modules
$V(\mu)$ appearing in the decomposition of
$V(\L_\a)\otimes V(\L_\b)$. There is an edge between two vertices $V(\mu)$
and $V(\mu')$ iff
\beq
V(\mu')\subset V_{adj}\otimes V(\mu) ~~~\mbox{ and }~~
\epsilon(\mu)\epsilon(\mu')=-1.\label{adj}
\eeq
\vskip.1in
The condition in \reff{adj} is a necessary condition for \reff{link}
\cite{Zha91}.
This means that every link contained in the TPG is contained
also in the ETPG but the latter may contain
more links. Only if the ETPG is a tree do we
know that it is equal to the TPG. If we impose a
relation (\ref{2rho2}) on the $\rho$'s
for every link in the ETPG, we may be
imposing too many relations and thus may not always find a solution.
If however we do
find a solution, then this is the unique correct solution which we
would have obtained also from the TPG.

We now apply the above formalism to the one-parameter family of irreps
of $U_q(gl(m|n))$, all irreps of which are known to be affinizable.

Choose $\{\varepsilon_i\}^m_{i=1}\bigcup \{\bar{\varepsilon}_j\}^n_{j=1}$ as a
basis for the dual of the Cartan subalgebra of $gl(m|n)$ satisfying
\begin{equation}
(\varepsilon_i,\varepsilon_j)=\delta_{ij},~~~~
(\bar{\varepsilon}_i,\bar{\varepsilon}_j)=-\delta_{ij},~~~~
(\varepsilon_i,\bar{\varepsilon}_j)=0\,.
\end{equation}
Using this basis, any weight $\Lambda$ may written as
\begin{equation}
\Lambda\equiv (\Lambda_1,\cdots,\Lambda_m|\bar{\Lambda}_1,
\cdots, \bar{\Lambda}_n)\equiv \sum_{i=1}^m\Lambda_i\varepsilon_i+
\sum_{j=1}^n\bar{\Lambda}_j\bar{\varepsilon}_j
\end{equation}
and the graded half sum $\rho$ of the positive roots of $gl(m|n)$ is
\begin{equation}
2\rho=\sum_{i=1}^m(m-n-2i+1)\varepsilon_i+\sum_{j=1}^n(m+n-2j+1)
\bar{\varepsilon}_j\,.
\end{equation}
We assume $m\geq n$ and for $0\leq k\leq mn$
we call a Young diagram
$[\l]=[\l_1,\l_2,\cdots,\l_r],~\l_1\geq\l_2\cdots\geq\l_r\geq 0$ for
the permutation group $S_k$ (i.e. $\l_1+\l_2+\cdots+\l_r=k$)
{\bf allowable}, if it has at most $n$ columns and $m$ rows; i.e.
$r\leq m,~\l_i\leq n$.
Associated with each such Young diagram $[\l]$
we define a weight of $gl(m|n)$
\beq
\L_{[\l]}=(\dot{0}_{m-r},-\l_r,\cdots,-\l_1|\underbrace{r,\cdots,r}_{\l_r},
  \underbrace{r-1,\cdots,r-1}_{\l_{r-1}-\l_r},\cdots,
  \underbrace{1,\cdots,1}_{\l_1-\l_2},
  \underbrace{0,\cdots,0}_{n-\l_1})\label{weight}
\eeq
In what follows we will consider the
one-parameter family of finite-dimensional irreducible
$U_q(gl(m|n))$-modules $V(\L_\alpha)$ with highest weights of the form
$\Lambda_\alpha=(0,\cdots,0|
\alpha,\cdots,\alpha)\equiv (\dot{0}|\dot{\a})$.
These irreps $V(\L_\a)$ are unitary of type I
if $\a>n-1$ and unitary of type II if $\a<1-m$. Here we assume real
$\a$ satisfying one of these conditions, in which case $V(\L_\a)$ is
also typical of dimension $2^{mn}$.

We have the following decomposition of $V(\L_\a)$
into irreps of the even subalgebra
$gl(m)\bigoplus gl(n)$:
\beq
V(\L_\a)=\bigoplus^{mn}_{k=0}\bigoplus_{[\l]\in S_k}'
  V_0(\L_{[\l]}+\L_\a)
\eeq
where the prime signifies the summation over allowed $k$-box Young
diagrams. Note that the index $k$ gives the ${\bf Z}$-graded level of
the irrep concerned. Alternatively we may simply write
\beq
V(\L_\a)=\bigoplus_{[\l]}'
  V_0(\L_{[\l]}+\L_\a)
\eeq
The number of boxes then gives the level. For $\L_\a,~\L_\b$ of same
type, we immediately deduce the tensor product decomposition
\beq
V(\L_\a)\otimes V(\L_\b)=\bigoplus'_{[\l]}
  V(\L_{[\l]}+\L_{\a+\b})\label{decom}
\eeq
The eigenvalue of the second order Casimir on the irrep $V(\L_{[\l]}+
\L_{\a+\b})$ can be shown to be
\beq\label{casimirg}
C([\l])=2\sum_{i=1}^r\l_i(\l_i+1-\a-\b-2i)-n(\a+\b)(\a+\b+m)\label{casimir}
\eeq
Below we show that the ETPG corresponding to the above tensor product
is always consistent and we derive an explicit expression for the
eigenvalues $\rho_{[\l]}(x)$ for the R-matrix $\check{R}(x|\a,\b)$ of
$U_q(gl(m|n))$.
It is instructive to first consider some examples.

\newsavebox{\yo}
\savebox{\yo}(4,8){\begin{picture}(8,8)(-2,0)
\put(0,0){\framebox(4,4){}}
\end{picture}}

\newsavebox{\yt}
\savebox{\yt}(4,8){\begin{picture}(8,11)(-2,0)
\multiput(0,0)(0,4){2}{\framebox(4,4){}}
\end{picture}}

\newsavebox{\yr}
\savebox{\yr}(4,10){\begin{picture}(8,12)(-2,3)
\multiput(0,0)(0,4){3}{\framebox(4,4){}}
\end{picture}}

\newsavebox{\yto}
\savebox{\yto}(10,8){\begin{picture}(10,11)(-2,0)
\multiput(0,0)(0,4){2}{\framebox(4,4){}}
\put(4,4){\framebox(4,4){}}
\end{picture}}

\newsavebox{\yro}
\savebox{\yro}(10,10){\begin{picture}(10,12)(-2,3)
\multiput(0,0)(0,4){3}{\framebox(4,4){}}
\put(4,8){\framebox(4,4){}}
\end{picture}}

\newsavebox{\yrt}
\savebox{\yrt}(10,10){\begin{picture}(10,12)(-2,3)
\multiput(0,0)(0,4){3}{\framebox(4,4){}}
\multiput(4,4)(0,4){2}{\framebox(4,4){}}
\end{picture}}

\newsavebox{\ytt}
\savebox{\ytt}(10,8){\begin{picture}(10,11)(-2,0)
\multiput(0,0)(0,4){2}{\framebox(4,4){}}
\multiput(4,0)(0,4){2}{\framebox(4,4){}}
\end{picture}}

\newsavebox{\yoo}
\savebox{\yoo}(10,8){\begin{picture}(10,8)(-2,0)
\put(0,0){\framebox(4,4){}}
\put(4,0){\framebox(4,4){}}
\end{picture}}

\vskip.1in
\noindent{\bf Example (1): $U_q(gl(m|1))$}

In this case we have the tensor product decomposition
\bea
V(\L_\a)\otimes V(\L_\b)&=&V(\l_{\a+\b})\oplus
V(\dot{0},-1|\a+\b+1)\\
&&\oplus V(\dot{0},-1,-1|\a+\b+2)\oplus \cdots
\oplus V(-\dot{1}|\a+\b+m).\nonumber
\eea
In terms of Young diagrams the ETPG is

{\unitlength 1mm
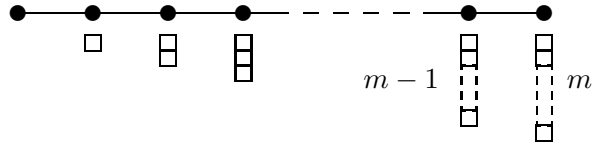
\begin{picture}(100,25)(-20,12)
\put(10,30){\line(1,0){36}}
\put(64,30){\line(1,0){16}}
\multiput(48,30)(4,0){4}{\line(1,0){2}}
\multiput(10,30)(10,0){4}{\circle*{2}}
\multiput(70,30)(10,0){2}{\circle*{2}}
\put(19,25){\framebox(2,2){}}
\multiput(29,25)(0,-2){2}{\framebox(2,2){}}
\multiput(39,25)(0,-2){3}{\framebox(2,2){}}
\multiput(69,25)(0,-2){2}{\framebox(2,2){}}
\multiput(79,25)(0,-2){2}{\framebox(2,2){}}
\put(69,17){\dashbox(2,6){}}
\put(69,15){\framebox(2,2){}}
\put(79,15){\dashbox(2,8){}}
\put(79,13){\framebox(2,2){}}
\put(56,20){$m-1$}
\put(83,20){$m$}
\end{picture}}

\noindent
which is obviously consistent.

\vskip.1in
\noindent{\bf Example (2): $U_q(gl(2|2))$}

The tensor product decomposition is
\bea
V(\L_\a)\otimes V(\L_\b)&=&V(\l_{\a+\b})\oplus
V({0},-1|\a+\b+1,\a+\b)\oplus
V(-1,-1|\a+\b+2,\a+\b)\nonumber\\
&&\oplus
V(0,-2|\a+\b+1,\a+\b+1)\oplus
V(-1,-2|\a+\b+2,\a+\b+1)\nonumber\\
&&\oplus
V(-2,-2|\a+\b+2,\a+\b+2).
\eea
The ETPG in terms of Young diagrams is shown in
figure 1a. From formula (\ref{casimirg}) for $C([\l])$ it is easily
deduced that
\beq
C(\usebox{\yto})-C(\usebox{\yt})=C(\usebox{\yoo})-C(\usebox{\yo})
=-2(\a+\b-2)
\eeq
so that this diagram is consistent. For completeness we note that
\bea
&&C(\cdot)=-2(\a+\b)(\a+\b+2),~~~~
C(\usebox{\yo})=-2(\a+\b)+C(\cdot),
\nonumber\\
&&C(\usebox{\yoo})=-4(\a+\b-1)+C(\cdot),~~~~
C(\usebox{\yt})=-4(\a+\b+1)+C(\cdot),
\nonumber\\
&&C(\usebox{\yto})=-6(\a+\b)+C(\cdot),~~~~
C(\usebox{\ytt})=-8(\a+\b)+C(\cdot).
\eea

\begin{figure}
\unitlength 1mm
\begin{picture}(100,65)(-30,0)
\put(20,50){\circle*{2}}
\put(20,40){\circle*{2}}
\put(12,32){\circle*{2}}
\put(28,32){\circle*{2}}
\put(20,24){\circle*{2}}
\put(20,14){\circle*{2}}
\put(20,40){\line(0,1){10}}
\put(20,40){\line(1,-1){8}}
\put(20,40){\line(-1,-1){8}}
\put(20,24){\line(0,-1){10}}
\put(20,24){\line(1,1){8}}
\put(20,24){\line(-1,1){8}}
\put(22,42){\framebox(2,2){}}
\multiput(6,31)(2,0){2}{\framebox(2,2){}}
\multiput(30,32)(0,2){2}{\framebox(2,2){}}
\multiput(12,22)(2,0){2}{\framebox(2,2){}}
\put(12,20){\framebox(2,2){}}
\multiput(23,12)(2,0){2}{\framebox(2,2){}}
\multiput(23,14)(2,0){2}{\framebox(2,2){}}
\put(16,5){(a)}

\put(70,60){\circle*{2}}
\put(70,50){\circle*{2}}
\put(62,42){\circle*{2}}
\put(78,42){\circle*{2}}
\put(70,34){\circle*{2}}
\put(62,26){\circle*{2}}
\put(78,26){\circle*{2}}
\put(70,18){\circle*{2}}
\put(70,8){\circle*{2}}
\put(86,34){\circle*{2}}
\put(70,50){\line(0,1){10}}
\put(70,50){\line(1,-1){8}}
\put(70,50){\line(-1,-1){8}}
\put(70,34){\line(1,-1){8}}
\put(70,34){\line(1,1){8}}
\put(70,34){\line(-1,1){8}}
\put(70,34){\line(-1,-1){8}}
\put(70,18){\line(-1,1){8}}
\put(70,18){\line(1,1){8}}
\put(70,18){\line(0,-1){10}}
\put(86,34){\line(-1,-1){8}}
\put(86,34){\line(-1,1){8}}
\put(72,52){\framebox(2,2){}}
\multiput(56,41)(2,0){2}{\framebox(2,2){}}
\multiput(80,42)(0,2){2}{\framebox(2,2){}}
\multiput(62,32)(2,0){2}{\framebox(2,2){}}
\put(62,30){\framebox(2,2){}}
\multiput(88,32)(0,2){3}{\framebox(2,2){}}
\multiput(81,20)(0,2){3}{\framebox(2,2){}}
\multiput(73,13)(0,2){3}{\framebox(2,2){}}
\multiput(72,4)(0,2){3}{\framebox(2,2){}}
\multiput(74,4)(0,2){3}{\framebox(2,2){}}
\multiput(75,15)(0,2){2}{\framebox(2,2){}}
\multiput(56,24)(0,2){2}{\framebox(2,2){}}
\multiput(58,24)(0,2){2}{\framebox(2,2){}}
\put(83,24){\framebox(2,2){}}
\put(66,0){(b)}
\end{picture}
\caption{The ETPGs for $V(\L_\a)
\otimes V(\L_\b)$ in (a): $U_q(gl(2|2))$ and (b): $U_q(gl(3|2))$.}
\end{figure}
\vskip.1in
\noindent{\bf Example (2): $U_q(gl(3|2))$}

The ETPG is given in figure 1b. In this case
formula (\ref{casimirg}) gives
\bea
C(\usebox{\yto})-C(\usebox{\yt})=C(\usebox{\yoo})-C(\usebox{\yo})
&=&-2(\a+\b-2),\nonumber\\
C(\usebox{\yro})-C(\usebox{\yr})=C(\usebox{\yto})-C(\usebox{\yt})
&=&-2(\a+\b-2),\nonumber\\
C(\usebox{\yrt})-C(\usebox{\yro})=C(\usebox{\ytt})-C(\usebox{\yto})
&=&-2(\a+\b).
\eea
So again all closed loops are consistent leading to a consistent graph.
\vskip.1in

Now we return to the general case $U_q(gl(m|n))$.
Corresponding to the tensor product decomposition (\ref{decom}), we note that
given an (allowable) Young diagram $[\l]$, the number of
boxes in $[\l]$ gives the level of the irrep $V([\l]+\L_{\a+\b})$
in the ETPG. We denote by $[\l+\D_r]$ the Young diagram obtained
from $[\l]$ by increasing row $r$ by one box leaving the remaining
rows unchanged.

Therefore in general we necessarily have closed loops of the form
\beq
{\unitlength 1mm
\begin{picture}(40,28)(10,0)
\put(24,8){\circle*{2}}
\put(24,20){\circle*{2}}
\put(18,14){\circle*{2}}
\put(30,14){\circle*{2}}
\put(18,14){\line(1,1){6}}
\put(18,14){\line(1,-1){6}}
\put(30,14){\line(-1,1){6}}
\put(30,14){\line(-1,-1){6}}
\put(24,6){\makebox(0,0)[t]{$[\l+\D_r+\D_k]$}}
\put(14,14){\makebox(0,0)[r]{$[\l+\D_r]$}}
\put(24,23){\makebox(0,0)[b]{$[\l]$}}
\put(34,14){\makebox(0,0)[l]{$[\l+\D_k]$}}
\end{picture}}
\eeq
and note that
\bea
C([\l+\D_r])-C([\l])&=&2(\l_r+1)(\l_r +2-\a-\b-2r)-
2\l_r(\l_r+1-\a-\b-2r)\nonumber\\
&=&2(2\l_r+2-\a-\b-2r).
\eea
The important point, which is easily seen, is that (for $r\neq k$)
\beq
C([\l+\D_r+\D_k])-c([\l])=\left(C([\l+\D_r])-C([\l])\right)
+\left(C([\l+\D_k])-C([\l])\right).
\eeq
Thus
\beq
C([\l+\D_r+\D_k])-C([\l+\D_k])=C([\l+\D_r)-C([\l])
\eeq
so that all such closed loops are consistent. This shows that all the
ETPG's are consistent.


We now recast (\ref{ansatz}) in the form,
with the help of the Young diagram notation,
\beq
\check{R}(x|\a,\b)=\sum_{[\l]}'\rho_{[\l]}(x){\bf P}_{[\l]}^{\a\b},
\eeq
where the prime signifies the summation over allowed Young diagrams, as
in the tensor product decomposition (\ref{decom}).
Since the ETPG is consistent we may calculate
the coefficients $\rho_{[\l]}(x)$ by succesive removal of boxes, starting
with the last column and proceeding to eliminate column by column. By
this means we arrive at
\beq
\rho_{[\l]}(x)=\prod_{l=1}^r\prod_{k=1}^{\l_l}
\langle 2k-\a-\b-2l\rangle,~~~~~~[\l]\equiv[\l_1,\l_2,\cdots\l_r]
\eeq
where we have chosen the normalization $\rho.(x)=1$ (corresponding to
the highest vertex $\L_{\a+\b}$) and the notation $\langle a\rangle$
is as in (\ref{angle}).

\vskip.3in
\begin{center}
{\bf Acknowledgements:}
\end{center}
We thank Anthony J. Bracken for collaborations. The financial
support from the Australian Research Council is gratefully acknowledged.

\newpage


\begin{thebibliography}{99}
\bibitem{DGZ} G.W.Delius, M.D.Gould and Y.-Z.Zhang, {\em On the construction
  of trigonometric solutions of the Yang-Baxter equation},
  hep-th/9405030, {\em Nucl.Phys.} {\bf B} (in press).
\bibitem{Bra94a} A.J.Bracken, G.W.Delius, M.D.Gould and Y.-Z.Zhang,
  {\em J.Phys.} {\bf A:} {\em Math.Gen.} {\bf 27} (1994) 6551.
\bibitem{Del94b} G.W.Delius, M.D.Gould, J.R.Links and Y.-Z.Zhang,
  {\em On type-I quantum affine superalgebras}, hep-th/9408006,
  submitted to {\em Int.J.Mod.Phys.} {\bf A}.
\bibitem{Bra94b} A.J.Bracken, M.D.Gould, J.R.Links and Y.-Z.Zhang,
  {\em A new supersymmetric and exactly solvable model of correlated
  electrons}, cond-mat/9410026.
\bibitem{Kac77} V.G.Kac, {\em Adv.Math.} {\bf 26} (1977) 8; and in
  {\em Lect. Notes in Math.} {\bf 676} (1978) 597.
\bibitem{MS93} M.D.Gould and M.Scheunert, {\em Classification of finite
  dimensional unitary irreps for $U_q(gl(m|n))$}, The University of
  Queensland preprint, 1993.
\bibitem{Baz90} V.V.Bazhanov, R.M.Kashaev, V.V.Mangazeev and Yu.Stroganov,
  {\em Commun.Math.Phys.} {\bf 138} (1991) 393.
\bibitem{Dat91} E.Date, M.Jimbo, K.Miki and T.Miwa,
  {\em Commun.Math.Phys.} {\bf 137} (1991) 133.
\bibitem{Jimbo} M.Jimbo, {\em Lett.Math.Phys.} {\bf 10} (1985) 63;
  {\bf 11} (1986) 247.
\bibitem{Res88} N.Reshetikhin, {\em Quantized universal enveloping
  algebras, the Yang-Baxter equation and invariants of links: I, II},
  preprints LOMI E-4-87, E-17-87 (1988).
\bibitem{Dri90} V.G.Drinfeld, {\em Leningrad Math.J.} {\bf 1} (1990) 321.
\bibitem{Zha91}  R.B.Zhang, M.D.Gould and A.J.Bracken, {\em Nucl.Phys.}
  {\bf B354} (1991) 625.


\end{thebibliography}
\end{document}